\documentclass[sigconf]{acmart}

\usepackage{booktabs}
\usepackage{graphicx}
\usepackage{amsmath}
\usepackage{adjustbox}
\usepackage{colortbl}
\usepackage{tabularx}
\usepackage{multirow}
\usepackage{booktabs}
\usepackage{algorithm2e}
\usepackage{balance}
\usepackage{url}
\usepackage{xcolor}

\usepackage{fancyvrb}

\copyrightyear{2025}
\acmYear{2025}
\acmDOI{XXXXXXX.XXXXXXX}
\acmConference[SAC'25]{The 40th ACM/SIGAPP Symposium On Applied Computing}{March 31--April 04, 2025}{Catania, Italy}
\acmISBN{979-X-XXXX-XXXX-X/26/03}

\acmArticle{4}

\begin{document}

\pagenumbering{gobble}
\setcounter{page}{0}

\section*{ACM Copyright Statement}

\noindent
Copyright \textcopyright 2025 held by the owner/author(s). Publication rights licensed to ACM. Permission to make digital or hard copies of all or part of this work for personal or classroom use is granted without fee provided that copies are not made or distributed for profit or commercial advantage and that copies bear this notice and the full citation on the first page. Copyrights for components of this work owned by others than the author(s) must be honored. Abstracting with credit is permitted. To copy otherwise, or republish, to post on servers or to redistribute to lists, requires prior specific permission and/or a fee. Request permissions from permissions@acm.org.

\vspace{3em}
\noindent\textbf{Published in:} Proceedings of the 40th ACM/SIGAPP Symposium On Applied Computing, 2025 (\textit{SAC'25, Catania, Italy})

\vspace{3em}
\noindent\textbf{Cite as:}
\vspace{0.5em}

\noindent
Pedro R. Pires, Rafael T. Sereicikas, Gregorio F. Azevedo, and Tiago A. Almeida. 2025. Collaborative Filtering Through Weighted Similarities of User and Item Embeddings. In \textit{Proceedings of the 40th ACM/SIGAPP Symposium On Applied Computing} (Catania, Italy) \textit{(SAC'25)}. Association for Computing Machinery, New York, NY, USA, 1188-1195, doi:10.1145/3672608.3707877

\vspace{3em}
\noindent\textbf{BibTeX:}
\vspace{0.5em}

\begin{Verbatim}[frame=single]
@inproceedings{10.1145/3672608.3707877,
    title       =   {Collaborative Filtering Through Weighted Similarities of User and Item Embeddings},
    author      =   {Pedro R. Pires and Rafael T. Sereicikas and Gregorio F. Azevedo and Tiago A. Almeida},
    booktitle   =   {Proceedings of the 40th ACM/SIGAPP Symposium On Applied Computing},
    series      =   {SAC'25},
    location    =   {Catania, Italy},
    publisher   =   {Association for Computing Machinery},
    address     =   {New York, NY, USA},
    pages       =   {1188--1195},
    year        =   {2025},
    doi         =   {10.1145/3672608.3707877},
    keywords    =   {Recommender systems, Collaborative filtering, Hybrid recommender, 
                     Distributed vector representation, Embeddings, Ensemble},
}
\end{Verbatim}

\newpage
\pagenumbering{arabic}

\title{Collaborative Filtering Through Weighted Similarities of User and Item Embeddings}
  
\renewcommand{\shorttitle}{CF Through Weighted Similarities of User and Item Embeddings}

\author{Pedro R. Pires}
    \authornote{Corresponding author}
    \orcid{0000-0001-7990-9097}
    \affiliation{
        \institution{Federal University of São Carlos}
        \streetaddress{Rod. João Leme dos Santos km 110}
        \city{Sorocaba} 
        \state{São Paulo} 
        \country{Brazil}
        \postcode{18052-780}  
    }
    \email{pedro.pires@dcomp.sor.ufscar.br}

\author{Rafael T. Sereicikas}
    \orcid{0009-0009-9198-5469}
    \affiliation{
        \institution{Federal University of São Carlos}
        \streetaddress{Rod. João Leme dos Santos km 110}
        \city{Sorocaba} 
        \state{São Paulo} 
        \country{Brazil}
        \postcode{18052-780}  
    }
    \email{rafaeltofoli@estudante.ufscar.br}

\author{Gregorio F. Azevedo}
    \orcid{0000-0002-1096-7456}
    \affiliation{
        \institution{Federal University of São Carlos}
        \streetaddress{Rod. João Leme dos Santos km 110}
        \city{Sorocaba} 
        \state{São Paulo} 
        \country{Brazil}
        \postcode{18052-780}  
    }
    \email{gregorio.fornetti@estudante.ufscar.br}

\author{Tiago A. Almeida}
    \orcid{0000-0001-6943-8033}
    \affiliation{
        \institution{Federal University of São Carlos}
        \streetaddress{Rod. João Leme dos Santos km 110}
        \city{Sorocaba} 
        \state{São Paulo} 
        \country{Brazil}
        \postcode{18052-780}  
    }
    \email{talmeida@ufscar.br}


\begin{abstract}
In recent years, neural networks and other complex models have dominated recommender systems, often setting new benchmarks for state-of-the-art performance. Yet, despite these advancements, award-winning research has demonstrated that traditional matrix factorization methods can remain competitive, offering simplicity and reduced computational overhead. Hybrid models, which combine matrix factorization with newer techniques, are increasingly employed to harness the strengths of multiple approaches. This paper proposes a novel ensemble method that unifies user-item and item-item recommendations through a weighted similarity framework to deliver top-$N$ recommendations. Our approach is distinctive in its use of shared user and item embeddings for both recommendation strategies, simplifying the architecture and enhancing computational efficiency. Extensive experiments across multiple datasets show that our method achieves competitive performance and is robust in varying scenarios that favor either user--item or item--item recommendations. Additionally, by eliminating the need for embedding-specific fine-tuning, our model allows for the seamless reuse of hyperparameters from the base algorithm without sacrificing performance. This results in a method that is both efficient and easy to implement. Our open-source implementation is available at \url{https://github.com/UFSCar-LaSID/weighted-sims-recommender}.
\end{abstract}

%
%

\ccsdesc[500]{Computing methodologies~Ensemble methods}
\ccsdesc[300]{Computing methodologies~Factorization methods}
\ccsdesc[100]{Computing methodologies~Neural networks}

\keywords{Recommender systems, Collaborative filtering, Hybrid recommender, Distributed vector representation, Embeddings, Ensemble.}

\maketitle

\section{Introduction}

Recommender systems (RS) have become indispensable in modern digital platforms, helping users discover relevant content and enhancing engagement~\cite{Bobadilla2013}. Since their inception in the mid-90s, initially designed for specialized domains such as e-mail and news filtering~\cite{Resnick1994}, RS have evolved significantly~\cite{Bobadilla2013}. Today, they are ubiquitous across various websites and digital platforms, representing a critical research area with substantial investments and revenue generation for technology companies~\cite{Hallinan2016}.

Early RS models represented users and items as consumption vectors, with dimensions corresponding to the number of users and items~\cite{Herlocker2002}. As RS technologies advanced and their usage increased, these vectors expanded dramatically, leading to high-dimensional and sparse data representations~\cite{Khusro2016}. The increased demand for memory and processing power rendered traditional algorithms inefficient or impractical for large-scale applications.

To address these challenges, researchers turned to embeddings, low-dimensional dense vectors encapsulating intrinsic meanings \cite{Koren2009a}. Embeddings are widely used in various artificial intelligence domains, including natural language processing~\cite{Mikolov2013} and computer vision~\cite{Dosovitskiy2020}, where they have consistently delivered impressive and efficient results. Embeddings in RS gained prominence during the Netflix Prize Challenge, with the advent of matrix factorization (MF) models~\cite{Koren2009a}. Singular value decomposition (SVD), an early method for collaborative filtering~\cite{Funk2006}, demonstrated promising results and inspired numerous subsequent algorithms in the competition, including the winning approach~\cite{Koren2009b}.

MF algorithms decompose a sparse interaction matrix into two smaller matrices, representing users and items in a low-dimensional vector space~\cite{Koren2009a}. The original matrix can be reconstructed by multiplying these smaller matrices, enabling the generation of recommendations. MF-based RS quickly became a prolific research topic, adapted for contexts such as implicit feedback~\cite{Hu2008} and enhanced by incorporating additional information~\cite{Rendle2010}.

With the rise of neural networks in machine learning, many neural network-based models have been proposed for recommendation tasks~\cite{Zhang2019}. Often more complex than traditional MF, these models aimed to learn distributed vector representations to compute recommendations~\cite{Barkan2016,Li2017,Ebesu2018,Hu2018}, achieving excellent results compared to baseline algorithms. However, recent discussions have questioned neural models' reliability and perceived superiority. Studies have shown that simple neighborhood or matrix factorization methods can outperform many neural network algorithms~\cite{Cremonesi2019}, highlighting a reproducibility crisis in the field~\cite{Lops2023} and renewing interest in simpler models~\cite{Pires2022}.

This paper introduces a novel recommendation ensemble method that combines user-item and item-item recommendation strategies, leveraging the strengths of both approaches~\cite{Deshpande2004}. Unlike previous methods that trained models independently before combining them, our approach uses pre-trained embeddings generated by MF models or neural networks. The final recommendations are derived by calculating similarities using these embeddings and weighting the two recommendation strategies. This method offers ease of application and low computational processing demands. To address the reproducibility crisis, all datasets used in this study are publicly available, and the implemented code is available in the following repository: \url{https://github.com/UFSCar-LaSID/weighted-sims-recommender}.

\section{Related work}

Matrix factorization (MF) remains one of the most prominent methods in recommender systems (RS)~\cite{Bobadilla2013}. By learning low-dimensional distributed vector representations of users and items~\cite{Koren2009a}, MF effectively addresses the challenges posed by high-dimensional, sparse interaction data~\cite{Khusro2016}. Since its success in the Netflix Prize~\cite{Funk2006}, numerous extensions have been proposed, such as leveraging implicit feedback~\cite{Salakhutdinov2007,Hu2008}, incorporating temporal dynamics~\cite{Koren2009b,Matuszyk2015}, employing novel optimization techniques~\cite{Rendle2009,Luo2014}, and integrating MF with other machine learning algorithms~\cite{Rendle2010,Lara2020}. These advancements have solidified MF as a cornerstone in collaborative filtering.

In recent years, the popularity of neural network-based methods has surged in various machine learning domains, including RS~\cite{Zhang2019}. Despite the promising results of many proposed models, recent studies draw attention to a reproducibility crisis in the area of RS~\cite{Lops2023}. Moreover, some studies have shown that complex methods considered state-of-the-art are easily outperformed by simpler neighborhood-based or MF-based methods~\cite{Cremonesi2019}. This has led to a renewed focus on simpler algorithms that demand much less computational power and are more efficient~\cite{Pires2022}.

One effective strategy for creating powerful and efficient algorithms is to combine multiple simpler algorithms into a recommender ensemble, also known as a hybrid recommender. Among the various strategies for creating hybrid RS, mixed ensembles combine RS modularly and are particularly effective. This approach allows for integrating recommender algorithms with complementary strengths or mitigating each other's weaknesses~\cite{Cano2017}.

Due to their popularity and proven efficiency, hybrid methods combining MF techniques are prevalent in the field~\cite{Cano2017}. The most common approach is to merge collaborative filtering algorithms based on MF with content-based algorithms~\cite{Suriati2017,Geetha2018,Parthasarathy2022}. In these ensembles, content-based recommendations help alleviate the cold-start problem by recommending items yet to be consumed~\cite{Cano2017}, addressing one of the main issues of collaborative filtering. However, the reliance on content information is a significant drawback because it may only sometimes be available.

Pure hybrid collaborative filtering algorithms often employ more advanced techniques, which can increase complexity and processing demands~\cite{Cano2017}. Nevertheless, many algorithms combine more straightforward techniques effectively. For instance, $\Gamma$UICF~\cite{Aljunid2021} combines user-based and item-based similarity algorithms, and Yu et al.~\cite{Zhang2022} calculate average similarities between users and items by selecting core items and users.

Building upon this foundation, our work introduces a novel ensemble model for top-$N$ recommendation within pure collaborative filtering, relying solely on the user-item interaction matrix. Inspired by recent findings that neighborhood-based methods often outperform deep learning models in RS tasks~\cite{Cremonesi2019}, our approach merges user-item and item-item similarity in a unified framework. Unlike existing hybrid methods that rely on independent training processes for each recommendation strategy, our model leverages shared embeddings generated from either neural networks or matrix factorization models. We streamline the process by employing the same embeddings for both user-item and item-item recommendations, requiring only a single embedding learning step. This reduces the computational burden and simplifies the architecture, offering a more efficient solution for generating high-quality recommendations.

\section{Weighted similarity of embeddings}\label{sec:weighted_sims}

Embedding-based algorithms represent users and items in a shared latent vector space. Typically, recommendations are generated by computing the dot product between user and item embeddings, identifying the most relevant items for a given user~\cite{Koren2009a}. On the other hand, item-based recommendations rely on items that a user has already interacted with, finding similar items based on their content or collaborative feedback~\cite{Ricci2011}. Our method combines the strengths of both approaches within a unified ensemble framework.

\begin{figure}[!htbp]
\centering
\includegraphics[width=0.45\textwidth]{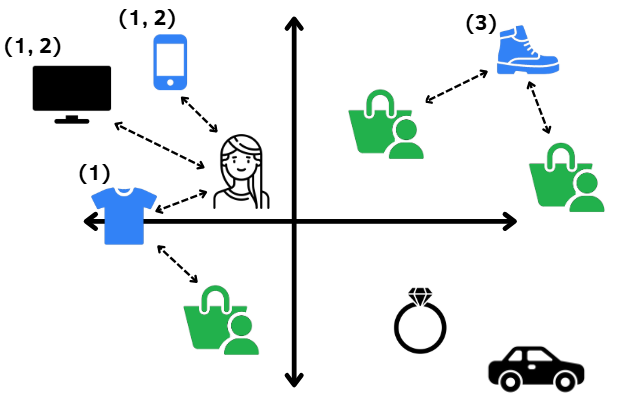}
\caption{Visual representation of using the weighted similarity of embeddings, with items consumed by the user represented in green, and similar items denoted by the dashed arrows. When using a traditional user--item similarity recommender, only items close to the user are selected (1), which may include items that are similar to the user but unrelated to their consumed items (2). When using the hybrid model with an item--item similarity, items distant from the user but close to their consumption history would also be recommended (3). The final recommendation would then be composed of items similar to both the user and their consumed items, as represented in blue.}
\Description[Visual representation of using the weighted similarity of embeddings]{Image of multiple icons scattered across a vector space representation, with two intersecting arrows in the center representing the axes. An icon of a woman, representing a user, is located at the center, with icons of items placed around her, representing both new items and items previously purchased by the woman. Dashed arrows connect the woman with items close to her and further with consumed items, representing cases where the system recommends items similar to the user or similar to items consumed by the user.}
\label{fig:vector_space}
\end{figure}

For item-based similarity, we leverage the same embeddings learned by the underlying embedding-based algorithms. This reuse of embeddings is the core efficiency advantage of our approach, allowing embeddings to be trained once and utilized for both user--item and item--item recommendations. By operating within a shared vector space, our method identifies items similar to the user’s preferences and also to items they have previously consumed. This broadens the recommendation search space while maintaining the semantic coherence captured by the embeddings, as shown in Figure~\ref{fig:vector_space}.

\begin{figure}[!htbp]
\centering
\includegraphics[width=0.48\textwidth]{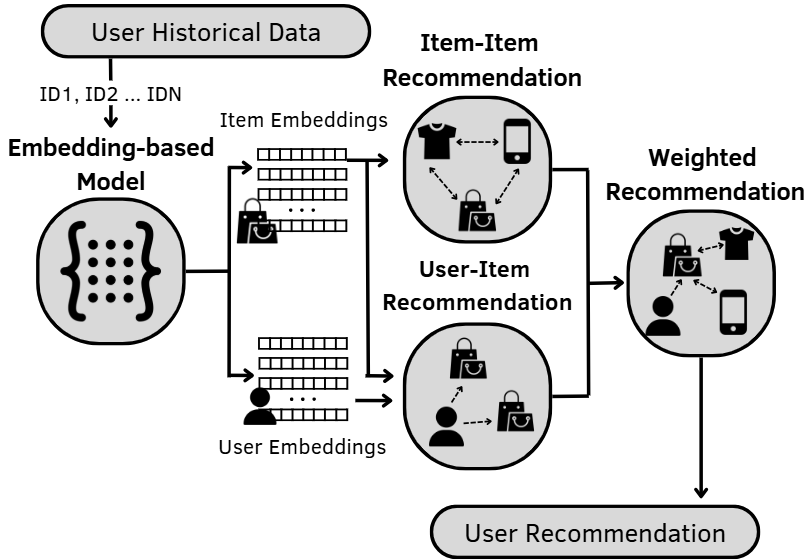}
\caption{Framework of the recommendation through weighted similarities of user and item embeddings.}
\Description[Diagram of the proposed algorithm]{Image of a diagram representing the pipeline of the weighted similarities algorithm, with steps represented as spherical shapes connected through arrows. The pipeline goes as follows: User Historical Data, Embedding-based Model, Item Embeddings and User Embeddings, Item-Item and User-Item Recommendation, Weighted Recommendation, and User Recommendation.}
\label{fig:weighted_sims}
\end{figure}

The proposed algorithm consists of four main steps: \textit{(i)} embedding generation using an external model, \textit{(ii)} calculation of user--item similarities, \textit{(iii)} calculation of item--item similarities, and \textit{(iv)} aggregation of these scores to produce the final recommendation. Unlike other ensemble methods that use different models for user--item and item--item similarities~\cite{Aljunid2021}, our method computes both using a single set of learned embeddings, simplifying the architecture. An illustration of the whole process is provided in Figure~\ref{fig:weighted_sims}.

\subsection{Embedding Generation}

The first step generates user and item embeddings via an external embedding model. We evaluated several models, including alternating least squares (ALS)~\cite{Hu2008}, Bayesian personalized ranking (BPR)~\cite{Rendle2009}, and RecVAE, a state-of-the-art variational autoencoder recommender~\cite{Shenbin2019}. These models are well-established and effective at capturing latent patterns in user--item interactions without requiring explicit feedback.

Once the embedding learning step is completed, we retrieve the user and item embeddings, denoted as $p$ (for users) and $q$ (for items). These embeddings are used in subsequent steps to generate user--item and item--item recommendations.

\subsection{User--Item Similarity Calculation}

We calculate the similarity between each user and all non-interacted items to generate user--item recommendations. This is performed using standard similarity metrics such as cosine similarity or dot-product. For a given user $u$, their embedding $p_u$ is compared with the embeddings of all unconsumed items $q_i$. The similarity score $R_{u,i}$ between user $u$ and item $i$ is computed as shown in Equation~\ref{eq:user_item_sim}:

\begin{equation}\label{eq:user_item_sim}
    R_{u,i} = p_u \cdot q_i
\end{equation}

This similarity score indicates how well item $i$ matches user $u$'s preferences based on their embedding representations.

\subsection{Item--Item Similarity Calculation}

In addition to user--item similarities, we compute item--item similarities using the same embeddings. For each user $u$, the set of items they have previously consumed, $I_u$, is retrieved. The similarity $S_{u,i}$ between a non-consumed item $i$ and the consumed items $j \in I_u$ is then calculated as the average similarity between $i$ and the items in $I_u$, as given in Equation~\ref{eq:item_item_sim}:

\begin{equation}\label{eq:item_item_sim}
    S_{u,i} = \frac{\sum_j^{I_u}{q_i \cdot q_j^T}}{|I_u|}
\end{equation}

This step captures how closely the new item $i$ relates to items the user has already interacted with, providing an alternative perspective to user--item similarity.

\subsection{Weighted Recommendation}

The final recommendation score $Z_{u,i}$ for item $i$ and user $u$ is calculated as a weighted combination of the user--item similarity $R_{u,i}$ and the item--item similarity $S_{u,i}$. The weights $w_R$ and $w_S$ control the relative contribution of each similarity algorithm, as described by Equation~\ref{eq:weighted_sim}:

\begin{equation}\label{eq:weighted_sim}
    \text{Z}_{u,i} = \frac{w_R \cdot R_{u,i} + w_S \cdot S_{u,i}}{w_R + w_S} \
\end{equation}

Once the scores for all non-consumed items are computed, the top-$N$ items with the highest scores are recommended to the user. The full algorithm is summarized in pseudocode in Algorithm~\ref{alg:weighted}. By integrating both user--item and item--item similarities in this way, our method is expected to provide a more balanced and enriched set of recommendations. 

\RestyleAlgo{ruled}

\begin{algorithm}[!htpb]
\caption{Weighted similarity algorithm}\label{alg:weighted}
\KwData{Set of users $U$, set of items $I$, user history $I_u$, users embeddings $p$,  items embeddings $q$, weights $w_R$ and $w_S$, recommendations list size $N$}
\KwResult{Top-$N$ recommendations list $rec_u$ for every user $u$}

$R \gets p \cdot q$ \tcp*{Store full user-item similarities}

\For{$u \in U$}{
    \For{$i \in I$}{
        \If{$i \notin I_u$}{
            $S_{u,i} \gets 0$\;
            \For{$j \in I_u$}{
                $S_{u,i} \gets S_{u,i} + \frac{q_i \cdot q_j^T}{|I_u|}$ \tcp*{Item similarities using user history}
            }
            $Z_{u,i} \gets \frac{w_R \cdot R_{u,i} \; + \; w_S \cdot S_{u,i}}{w_R \; + \; w_S} $ \tcp*{Compute final similarity between $u$ and $i$}
        }
    }

    \For{$n = 1$ to $N$}{
    $rec_{u,n} \gets \underset{i \in I; i \notin (rec_u \cup I_u)}{\text{argmax}} \, Z_{u,i}$ \tcp*{Retrieve the $n$-th most similar item for user $u$}
}
}
\end{algorithm}

\section{Experiments and results}\label{resultados}

This section outlines the experimental protocol for evaluating our hybrid model in a top-$N$ recommendation task. We first describe the datasets and preprocessing steps, followed by a discussion of the evaluation metrics and comparison with base recommenders. Our method's performance is demonstrated using two matrix factorization (MF) techniques and a state-of-the-art neural network method.

\subsection{Datasets and preprocessing}

The experiments were conducted on a diverse set of publicly available datasets commonly used in recommender systems research, as summarized in Table~\ref{tab:dados}. These datasets span multiple domains, and the columns $|U|$, $|I|$, and $|R|$ represent the number of users, items, and interactions, respectively, while $S$ denotes the sparsity rate, calculated as $S = 1 - \frac{|R|}{|U| \times |I|}$.

\begin{table}[!htpb]
\centering
\caption{Datasets used in the experiments.}
\label{tab:dados}
\begin{tabular}{lcccc}
\toprule
\textbf{Dataset} & $\boldsymbol{|U|}$ & $\boldsymbol{|I|}$ & $\boldsymbol{|R|}$ & $\boldsymbol{S}$\\
\midrule
\textbf{Anime}\footnotemark[1]          &    73,475 &  10,856 &  7,345,380 &   99.08\% \\
\textbf{BestBuy}\footnotemark[2]        &  1,268,702 &  69,858 &  1,862,782 &  99.99\% \\
\textbf{CiaoDVD}\footnotemark[3]        &    16,923 &  14,800 &    65,038 &   99.97\% \\
\textbf{Delicious}\footnotemark[4]      &     1,867 &  69,223 &   104,799 &   99.92\%  \\
\textbf{Filmtrust}\footnotemark[3]      &     1,492 &   1,881 &    28,579 &   98.98\% \\
\textbf{Jester}\footnotemark[5]         &   122,293 &    150 &  3,587,186 &   80.44\%  \\
\textbf{Last.FM} \footnotemark[4]       &     1,892 &  17,632 &    92,834 &   99.72\%  \\
\textbf{MovieLens-1M}\footnotemark[4]   &     6,039 &   3,628 &   836,478 &   96.18\% \\
\textbf{RetailRocket}\footnotemark[6]   &    11,719 &  12,025 &    21,270 &   99.98\%  \\
\bottomrule
\end{tabular}
\end{table}

\footnotetext[1]{Anime Recommendations dataset. Available at: \url{www.kaggle.com/CooperUnion/anime-recommendations-database} (visited on \today)}
\footnotetext[2]{Data Mining Hackathon on BIG DATA (7GB). Available at: \url{www.kaggle.com/c/acm-sf-chapter-hackathon-big} (visited on \today)}
\footnotetext[3]{FilmTrust and CiaoDVD. Available at: \url{https://guoguibing.github.io/librec/datasets.html} (visited on \today)}
\footnotetext[4]{GroupLens - Datasets. Available at: \url{www.grouplens.org/datasets/}} 
\footnotetext[5]{Jester dataset. Available at: \url{https://eigentaste.berkeley.edu/dataset/} (visited on \today)}
\footnotetext[6]{Retailrocket recommender system dataset. Available at \url{www.kaggle.com/retailrocket/ecommerce-dataset} (visited on \today)}

During preprocessing, duplicate interactions were removed, ensuring only one occurrence per user--item pair was retained. Additionally, inconsistent interactions (i.e., cases where the same user--item pair had different ratings) were excluded. For datasets with multiple interaction levels, we selected the one that best indicated user engagement, such as ``listen'' for Last.FM and ``buy'' for RetailRocket. For datasets with explicit feedback, we considered valid interactions only those with ratings higher than the intermediary rating, which can be calculated as shown in Equation~\ref{eq:intermediary_rating} considering $R$ as all the users' ratings of a specific dataset.

\begin{equation}\label{eq:intermediary_rating}
    min(R) + \frac{max(R) - min(R)}{2}
\end{equation}

\subsection{Experimental protocol}

The evaluation of our model followed a grid-search cross-validation approach. We randomly partitioned each dataset into five equal folds, using a five-fold cross-validation scheme to assess the model's performance. In each iteration, one fold was the test set, while the remaining four formed the training set. This ensured that each fold was used as a test set exactly once. To avoid cold-start issues, test interactions involving users or items absent from the training set were excluded, thus evaluating the models only on previously observed users and items.

For each fold, embeddings were learned from the user interactions in the training set. The recommender algorithms then used these embeddings to rank items, recommending the top-20 items for each user based on their interactions in the training set. The quality of the recommendations was evaluated by comparing the recommended items with those the user consumed in the test set.

Hyperparameter tuning was performed using grid search, optimizing for the Normalized Discounted Cumulative Gain at rank 10 (NDCG@10). We selected the hyperparameters that yielded the highest average NDCG@10 across the five folds as the optimal configuration for each algorithm.

\subsection{Embedding algorithms and parameter settings}

We employed two matrix factorization techniques optimized for implicit feedback: alternating least squares (ALS)~\cite{Hu2008} and Bayesian personalized ranking (BPR)\cite{Rendle2009}, alongside the state-of-the-art variational autoencoder model RecVAE~\cite{Shenbin2019}. All algorithms were implemented in \texttt{Python3}, with \texttt{Implicit}~\footnote{Implicit: Fast Python Collaborative Filtering for Implicit Datasets. Available at: \url{https://github.com/benfred/implicit} (visited on \today)} used for the MF-based methods and the original codebase for RecVAE.

For ALS, we tuned the number of epochs $({15, 30, 50})$, regularization factor $\lambda$ (${10^{-3}, 10^{-2}, 10^{-1}}$), and embedding size $({32, 64, 128})$. BPR was similarly tuned for epochs $({15, 30, 50})$, learning rate $\alpha$ (${10^{-3}, 10^{-2}, 10^{-1}}$), embedding size $({32, 64, 128})$, and regularization factor $\lambda$ (${10^{-3}, 10^{-2}, 10^{-1}}$). RecVAE was evaluated using $\gamma$ values from $({0.0035, 0.005, 0.01})$, with the remaining parameters set as per the original paper~\cite{Shenbin2019}.

\subsection{Recommendation and ensemble strategy}

After optimizing the hyperparameters, we selected the best-performing embeddings based on the NDCG@10 score for user--item recommendations. These embeddings were then used for both item--item similarity and the proposed hybrid ensemble. The item--item similarity algorithm required no additional tuning, as it directly leveraged the quality of the generated embeddings.

For the ensemble method, we tested two similarity metrics: dot-product (the default similarity for the embedding algorithms) and cosine similarity, a standard metric in recommender systems. To balance user--item and item--item recommendations, we evaluated various weight configurations, $w_R$ and $w_S$, corresponding to user--item and item--item scores, respectively. Weight ratios of \{1:4, 2:3, 1:1, 3:2, 4:1\} were tested to explore cases where either $w_R$ or $w_S$ dominated the recommendation process.

\begin{table*}[!htpb]
\centering
\caption{Hit-Rate in a top-10 recommendation task (HR@10).}
\label{tab:hitrate}
\begin{adjustbox}{width=.7\textwidth,center}
\begin{tabular}{lccc|ccc|ccc}
\toprule
\multirow{2}{*}{Dataset} & \multicolumn{3}{c}{ALS} & \multicolumn{3}{c}{BPR} & \multicolumn{3}{c}{RecVAE} \\
 & User-Item & Item-Item & Weighted & User-Item & Item-Item & Weighted & User-Item & Item-Item & Weighted \\
\midrule

Anime & \cellcolor[gray]{0.5}0.8376 & \cellcolor[gray]{0.95}0.5837 & \cellcolor[gray]{0.5312}0.8200 & \cellcolor[gray]{0.5}0.7924 & \cellcolor[gray]{0.95}0.5225 & \cellcolor[gray]{0.5415}0.7675 & \cellcolor[gray]{0.5122}0.9076 & \cellcolor[gray]{0.95}0.8248 & \cellcolor[gray]{0.5}\textbf{0.9099} \\

BestBuy & \cellcolor[gray]{0.6546}0.1258 & \cellcolor[gray]{0.95}0.0551 & \cellcolor[gray]{0.5}0.1628 & \cellcolor[gray]{0.95}0.2198 & \cellcolor[gray]{0.7713}0.2358 & \cellcolor[gray]{0.5}0.2601 & \cellcolor[gray]{0.5412}0.3674 & \cellcolor[gray]{0.95}0.3218 & \cellcolor[gray]{0.5}\textbf{0.3720} \\

CiaoDVD & \cellcolor[gray]{0.5}0.1185 & \cellcolor[gray]{0.95}0.0443 & \cellcolor[gray]{0.5164}0.1158 & \cellcolor[gray]{0.6032}0.0999 & \cellcolor[gray]{0.95}0.0515 & \cellcolor[gray]{0.5}0.1143 & \cellcolor[gray]{0.5}\textbf{0.1388} & \cellcolor[gray]{0.95}0.0713 & \cellcolor[gray]{0.5167}0.1363 \\

Delicious & \cellcolor[gray]{0.95}0.0964 & \cellcolor[gray]{0.5}0.4842 & \cellcolor[gray]{0.5198}0.4671 & \cellcolor[gray]{0.95}0.5003 & \cellcolor[gray]{0.5358}0.5211 & \cellcolor[gray]{0.5}\textbf{0.5229} & \cellcolor[gray]{0.95}0.4759 & \cellcolor[gray]{0.8384}0.4789 & \cellcolor[gray]{0.5}0.4880 \\

Filmtrust & \cellcolor[gray]{0.6556}0.5774 & \cellcolor[gray]{0.95}0.3060 & \cellcolor[gray]{0.5}0.7208 & \cellcolor[gray]{0.5}0.7713 & \cellcolor[gray]{0.95}0.6647 & \cellcolor[gray]{0.5321}0.7637 & \cellcolor[gray]{0.5014}0.8500 & \cellcolor[gray]{0.95}0.7196 & \cellcolor[gray]{0.5}\textbf{0.8504} \\

Jester & \cellcolor[gray]{0.5167}0.6995 & \cellcolor[gray]{0.95}0.6189 & \cellcolor[gray]{0.5}0.7026 & \cellcolor[gray]{0.511}0.8712 & \cellcolor[gray]{0.95}0.8313 & \cellcolor[gray]{0.5}0.8722 & \cellcolor[gray]{0.5053}0.9289 & \cellcolor[gray]{0.95}0.8784 & \cellcolor[gray]{0.5}\textbf{0.9295} \\

Last.FM & \cellcolor[gray]{0.5}\textbf{0.7237} & \cellcolor[gray]{0.95}0.5332 & \cellcolor[gray]{0.5347}0.7090 & \cellcolor[gray]{0.5109}0.6961 & \cellcolor[gray]{0.95}0.2435 & \cellcolor[gray]{0.5}0.7073 & \cellcolor[gray]{0.5}0.7087 & \cellcolor[gray]{0.95}0.4108 & \cellcolor[gray]{0.5003}0.7085 \\

MovieLens-1M & \cellcolor[gray]{0.5}0.8810 & \cellcolor[gray]{0.95}0.7116 & \cellcolor[gray]{0.5303}0.8696 & \cellcolor[gray]{0.5}0.7814 & \cellcolor[gray]{0.95}0.5607 & \cellcolor[gray]{0.5104}0.7763 & \cellcolor[gray]{0.5}\textbf{0.8830} & \cellcolor[gray]{0.95}0.7876 & \cellcolor[gray]{0.5}\textbf{0.8830} \\

RetailRocket & \cellcolor[gray]{0.77}0.2096 & \cellcolor[gray]{0.5}0.2108 & \cellcolor[gray]{0.95}0.2088 & \cellcolor[gray]{0.95}0.1961 & \cellcolor[gray]{0.6565}0.2051 & \cellcolor[gray]{0.5}0.2099 & \cellcolor[gray]{0.5}\textbf{0.2441} & \cellcolor[gray]{0.95}0.1015 & \cellcolor[gray]{0.5275}0.2354 \\
\bottomrule
\end{tabular}
\end{adjustbox}
\end{table*}

\begin{table*}[!htpb]
\centering
\caption{Normalized Discounted Cumulative Gain in a top-10 recommendation task (NDCG@10).}
\label{tab:ndcg}
\begin{adjustbox}{width=.7\textwidth,center}
\begin{tabular}{lccc|ccc|ccc}
\toprule
\multirow{2}{*}{Dataset} & \multicolumn{3}{c}{ALS} & \multicolumn{3}{c}{BPR} & \multicolumn{3}{c}{RecVAE} \\
 & User-Item & Item-Item & Weighted & User-Item & Item-Item & Weighted & User-Item & Item-Item & Weighted \\
\midrule

Anime & \cellcolor[gray]{0.5}0.4433 & \cellcolor[gray]{0.95}0.1617 & \cellcolor[gray]{0.5556}0.4085 & \cellcolor[gray]{0.5}0.3109 & \cellcolor[gray]{0.95}0.1240 & \cellcolor[gray]{0.52}0.3026 & \cellcolor[gray]{0.5343}0.5124 & \cellcolor[gray]{0.95}0.3369 & \cellcolor[gray]{0.5}\textbf{0.5269} \\

BestBuy & \cellcolor[gray]{0.6529}0.0690 & \cellcolor[gray]{0.95}0.0247 & \cellcolor[gray]{0.5}0.0918 & \cellcolor[gray]{0.95}0.1126 & \cellcolor[gray]{0.8217}0.1201 & \cellcolor[gray]{0.5}0.1389 & \cellcolor[gray]{0.5453}0.2014 & \cellcolor[gray]{0.95}0.1719 & \cellcolor[gray]{0.5}\textbf{0.2047} \\

CiaoDVD & \cellcolor[gray]{0.5}0.0444 & \cellcolor[gray]{0.95}0.0163 & \cellcolor[gray]{0.5192}0.0432 & \cellcolor[gray]{0.599}0.0381 & \cellcolor[gray]{0.95}0.0186 & \cellcolor[gray]{0.5}0.0436 & \cellcolor[gray]{0.5}\textbf{0.0533} & \cellcolor[gray]{0.95}0.0278 & \cellcolor[gray]{0.5194}0.0522 \\

Delicious & \cellcolor[gray]{0.95}0.0371 & \cellcolor[gray]{0.5}0.3928 & \cellcolor[gray]{0.5024}0.3909 & \cellcolor[gray]{0.95}0.4143 & \cellcolor[gray]{0.5295}0.4371 & \cellcolor[gray]{0.5}\textbf{0.4387} & \cellcolor[gray]{0.95}0.3211 & \cellcolor[gray]{0.51}0.3607 & \cellcolor[gray]{0.5}0.3616 \\

Filmtrust & \cellcolor[gray]{0.6869}0.2875 & \cellcolor[gray]{0.95}0.1556 & \cellcolor[gray]{0.5}0.3812 & \cellcolor[gray]{0.5}0.4568 & \cellcolor[gray]{0.95}0.3188 & \cellcolor[gray]{0.5157}0.4520 & \cellcolor[gray]{0.5}\textbf{0.5061} & \cellcolor[gray]{0.95}0.3801 & \cellcolor[gray]{0.5}\textbf{0.5061} \\

Jester & \cellcolor[gray]{0.5038}0.3625 & \cellcolor[gray]{0.95}0.2680 & \cellcolor[gray]{0.5}0.3633 & \cellcolor[gray]{0.5502}0.4571 & \cellcolor[gray]{0.95}0.3791 & \cellcolor[gray]{0.5}0.4669 & \cellcolor[gray]{0.5}\textbf{0.5592} & \cellcolor[gray]{0.95}0.4341 & \cellcolor[gray]{0.5309}0.5506 \\

Last.FM & \cellcolor[gray]{0.5039}0.2192 & \cellcolor[gray]{0.95}0.1173 & \cellcolor[gray]{0.5}\textbf{0.2201} & \cellcolor[gray]{0.506}0.1982 & \cellcolor[gray]{0.95}0.0366 & \cellcolor[gray]{0.5}0.2004 & \cellcolor[gray]{0.5}0.2131 & \cellcolor[gray]{0.95}0.0700 & \cellcolor[gray]{0.5003}0.2130 \\

MovieLens-1M & \cellcolor[gray]{0.5}\textbf{0.3917} & \cellcolor[gray]{0.95}0.1812 & \cellcolor[gray]{0.5066}0.3886 & \cellcolor[gray]{0.5}0.2512 & \cellcolor[gray]{0.95}0.1081 & \cellcolor[gray]{0.5145}0.2466 & \cellcolor[gray]{0.5049}0.3711 & \cellcolor[gray]{0.95}0.2714 & \cellcolor[gray]{0.5}0.3722 \\

RetailRocket & \cellcolor[gray]{0.95}0.0945 & \cellcolor[gray]{0.5875}0.1003 & \cellcolor[gray]{0.5}0.1017 & \cellcolor[gray]{0.95}0.0999 & \cellcolor[gray]{0.6595}0.1050 & \cellcolor[gray]{0.5}0.1078 & \cellcolor[gray]{0.5}\textbf{0.1281} & \cellcolor[gray]{0.95}0.0440 & \cellcolor[gray]{0.5096}0.1263 \\
\bottomrule
\end{tabular}
\end{adjustbox}
\end{table*}

\subsection{Results}

To evaluate the effectiveness of our proposed model of weighted similarities, we applied it across multiple embedding-based algorithms and compared its performance with the original recommender, which uses user--item similarities, and an item--item recommender with the same embeddings.

For each algorithm and dataset, we generated top-$N$ recommendations with $N$ ranging from 1 to 20. We computed the average Hit Rate and NDCG over five test folds. The Hit Rate, defined in Equation~\ref{eq:hit_rate}, measures the proportion of correctly recommended items, whereas NDCG evaluates the ranking quality of the recommendations, considering both the correctness and the position of items in the ranking (see Equations~\ref{eq:ndcg}, \ref{eq:idcg}, and \ref{eq:dcg}). Higher values of both metrics indicate better performance.

\noindent\begin{minipage}{.5\linewidth}
    \begin{equation}\label{eq:hit_rate}
        \text{HR} = \frac{\#\text{\textit{hits}}}{\#\text{\textit{users}}}
    \end{equation}
\end{minipage}%
\begin{minipage}{.5\linewidth}
    \begin{equation}\label{eq:ndcg}
        \text{NDCG} = \frac{DCG}{IDCG}
    \end{equation}
\end{minipage}

\begin{equation}\label{eq:idcg}
    \text{IDCG} = \sum_{u}^{U}\sum_{n=1}^{|I_u|}{\frac{1}{\log_2(n+1)}}
\end{equation}

\begin{equation}\label{eq:dcg}
    \text{DCG} = \sum_{u}^{U}\sum_{n=1}^{|rec_u|}{\frac{\text{\textit{rel}}}{\log_2(n+1)}}, \; \text{\textit{rel}} = 
        \begin{cases}
            1, & \text{if } \; rec_{u,n} \; \text{ is \textit{hit}}\\
            0, & \text{otherwise}
        \end{cases}
\end{equation}

Tables~\ref{tab:hitrate} and \ref{tab:ndcg} present the results for top-$10$ recommendations. Each column corresponds to a different recommender system and its underlying embedding algorithm. The columns labeled ``User-Item'' and ``Item-Item'' show the results for the baseline recommenders, while the ``Weighted'' column reflects the performance of our proposed model. Darker cells indicate superior performance when comparing the three types of recommenders for a given combination of dataset and embedding-based algorithm. The best overall results across every recommender and embedding-based algorithm are highlighted in \textbf{bold}.

Our ensemble model consistently delivers competitive performance across all metrics. In cases where either the user--item or item--item strategy outperforms the other, our model adapts to leverage their strengths. Overall, the ensemble model achieved the best results in 59\% of the cases, ranking as either the top or second-best model in nearly all scenarios except for Hit-Rate@10 on the RetailRocket dataset.

\begin{figure*}[!htpb]
\centering
\includegraphics[width=0.8\linewidth]{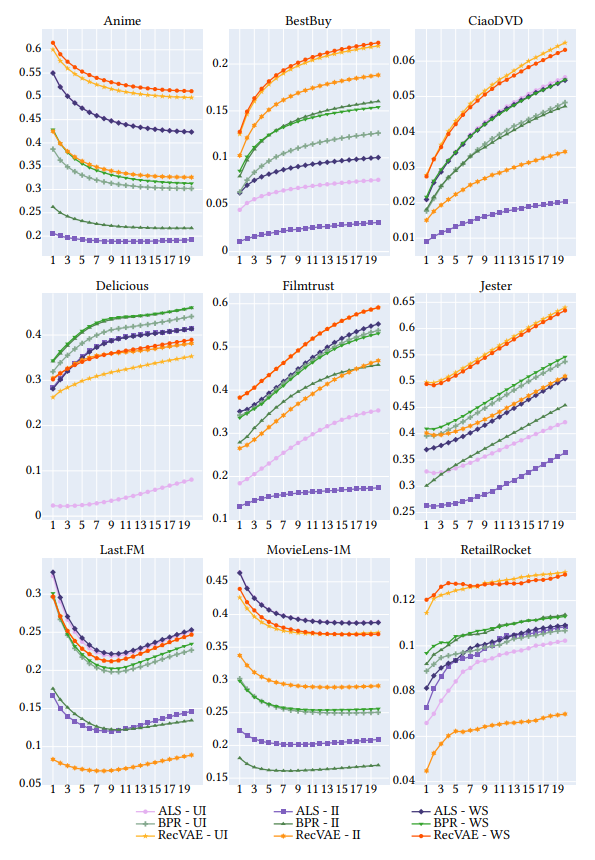}
\caption{NDCG@$N$ with $N$ ranging from 1 to 20. Recommender UI corresponds to user--item, II to item--item, and WS to the weighted similarities ensemble. For ease of comparison, each embedding-based algorithm follows a similar color palette.}
\Description[Graphs of the NDCG obtained by each algorithm in a top-N recommendation task for multiple values of N]{Grid of graphs with 3 rows and 3 columns. Each graph corresponds to a dataset and contains nine lines representing the 3 recommendation algorithms, i.e., user-item, item-item, and weighted similarity, for each of the three embedding-based algorithms, ALS, BPR, and RecVAE. Results vary according to each dataset, with specific embedding-based algorithms behaving better depending on the dataset. In general, the weighted similarity algorithm tends to be a superior method, being better or very similar to its counterparts.}
\label{fig:weightd_sims_ndcg}
\end{figure*}

Figure~\ref{fig:weightd_sims_ndcg} shows the performance of each recommender for varying values of $N$. Results demonstrate the stability of our approach. While the ensemble method occasionally underperforms its base algorithms at higher values of $N$, as observed for RecVAE and BPR on RetailRocket, and BPR on BestBuy, its performance remains robust overall.

Initially, we optimized the ensemble solely by tuning the weighting of the two recommendation strategies, ensuring adaptability across different datasets. This approach harnesses the strengths of both strategies, often outperforming the individual recommenders. Notably, our model achieves high efficiency by reusing embeddings tuned for user--item similarity, which eliminates the need for additional fine-tuning without significantly sacrificing performance. However, to fully explore the potential gains of fine-tuning, we re-evaluated the recommendations using the best-performing embeddings from each similarity algorithm. Table~\ref{tab:ndcg_finetuned} reports the NDCG@10 results after fine-tuning, with the percentage improvement compared to Table~\ref{tab:ndcg} shown in parentheses.

Fine-tuning significantly enhances performance, with improvements up to 21.3\% for the ensemble method. Interestingly, the gains are more pronounced for the item--item recommender, with an average improvement of 22.64\%, compared to 2.45\% for the ensemble. This suggests that while fine-tuning is beneficial, it is less critical for our ensemble model.

\begin{table*}[!htpb]
\centering
\caption{NDCG@10 and the relative improvements when performing fine-tuning of the embeddings.}
\label{tab:ndcg_finetuned}
\begin{adjustbox}{width=.95\textwidth,center}
\begin{tabular}{lccc|ccc|ccc}
\toprule
\multirow{2}{*}{Dataset} & \multicolumn{3}{c}{ALS} & \multicolumn{3}{c}{BPR} & \multicolumn{3}{c}{RecVAE} \\
 & User-Item & Item-Item & Weighted & User-Item & Item-Item & Weighted & User-Item & Item-Item & Weighted \\
\midrule

Anime & \cellcolor[gray]{0.5}0.4433 & \cellcolor[gray]{0.95}0.1887 (+16.7\%) & \cellcolor[gray]{0.5009}0.4428 (+8.4\%) & \cellcolor[gray]{0.5706}0.3109 & \cellcolor[gray]{0.95}0.2206 (+77.9\%) & \cellcolor[gray]{0.5}0.3277 (+8.3\%) & \cellcolor[gray]{0.5343}0.5124 & \cellcolor[gray]{0.95}0.3369 (+0.0\%) & \cellcolor[gray]{0.5}\textbf{0.5269} (+0.0\%) \\

BestBuy & \cellcolor[gray]{0.6547}0.0690 & \cellcolor[gray]{0.95}0.0249 (+0.8\%) & \cellcolor[gray]{0.5}0.0921 (+0.3\%) & \cellcolor[gray]{0.95}0.1126 & \cellcolor[gray]{0.5}0.1439 (+19.8\%) & \cellcolor[gray]{0.5403}0.1411 (+1.6\%) & \cellcolor[gray]{0.5453}0.2014 & \cellcolor[gray]{0.95}0.1719 (+0.0\%)  & \cellcolor[gray]{0.5}\textbf{0.2047} (+0.0\%) \\

CiaoDVD & \cellcolor[gray]{0.5}0.0444 & \cellcolor[gray]{0.95}0.0166 (+2.0\%) & \cellcolor[gray]{0.5097}0.0438 (+1.5\%) & \cellcolor[gray]{0.875}0.0381 & \cellcolor[gray]{0.95}0.0370 (+98.7\%) & \cellcolor[gray]{0.5}0.0436 (+0.0\%) & \cellcolor[gray]{0.5}\textbf{0.0533} & \cellcolor[gray]{0.95}0.0278 (+0.0\%) & \cellcolor[gray]{0.5194}0.0522 (+0.0\%) \\

Delicious & \cellcolor[gray]{0.95}0.0371 & \cellcolor[gray]{0.5}0.3932 (+0.1\%) & \cellcolor[gray]{0.5019}0.3917 (+0.2\%) & \cellcolor[gray]{0.95}0.4143 & \cellcolor[gray]{0.5445}0.4371 (+0.0\%) & \cellcolor[gray]{0.5}\textbf{0.4396} (+0.2\%) & \cellcolor[gray]{0.95}0.3211 & \cellcolor[gray]{0.5247}0.3607 (+0.0\%) & \cellcolor[gray]{0.5}0.3630 (+0.4\%) \\

Filmtrust & \cellcolor[gray]{0.7633}0.2875 & \cellcolor[gray]{0.95}0.1635 (+5.1\%) & \cellcolor[gray]{0.5}0.4624 (+21.3\%) & \cellcolor[gray]{0.5}0.4568 & \cellcolor[gray]{0.95}0.4058 (+27.3\%) & \cellcolor[gray]{0.5424}0.4520 (+0.0\%) & \cellcolor[gray]{0.5}\textbf{0.5061} & \cellcolor[gray]{0.95}0.3801 (+0.0\%) & \cellcolor[gray]{0.5}\textbf{0.5061} (+0.0\%) \\

Jester & \cellcolor[gray]{0.7068}0.3625 & \cellcolor[gray]{0.95}0.2902 (+8.3\%)  & \cellcolor[gray]{0.5}0.4240 (+16.7\%) & \cellcolor[gray]{0.5502}0.4571 & \cellcolor[gray]{0.95}0.3791 (+0.0\%) & \cellcolor[gray]{0.5}0.4669 (+0.0\%) & \cellcolor[gray]{0.5}\textbf{0.5592} & \cellcolor[gray]{0.95}0.4341 (+0.0\%) & \cellcolor[gray]{0.5209}0.5534 (+0.5\%) \\

Last.FM & \cellcolor[gray]{0.5121}0.2192 & \cellcolor[gray]{0.95}0.1214 (+3.5\%) & \cellcolor[gray]{0.5}\textbf{0.2219} (+0.8\%) & \cellcolor[gray]{0.5246}0.1982 & \cellcolor[gray]{0.95}0.1222 (+233.8\%)  & \cellcolor[gray]{0.5}0.2026 (+1.1\%) & \cellcolor[gray]{0.5}0.2131 & \cellcolor[gray]{0.95}0.0700 (+0.0\%) & \cellcolor[gray]{0.5003}0.2130 (+0.0\%) \\

MovieLens-1M & \cellcolor[gray]{0.5}\textbf{0.3917} & \cellcolor[gray]{0.95}0.2017 (+11.3\%) & \cellcolor[gray]{0.5019}0.3909 (+0.6\%) & \cellcolor[gray]{0.516}0.2512 & \cellcolor[gray]{0.95}0.1618 (+49.7\%) & \cellcolor[gray]{0.5}0.2545 (+3.2\%) & \cellcolor[gray]{0.5119}0.3711 & \cellcolor[gray]{0.95}0.2899 (+6.8\%) & \cellcolor[gray]{0.5}0.3733 (+0.3\%) \\

RetailRocket & \cellcolor[gray]{0.95}0.0945 & \cellcolor[gray]{0.5875}0.1003 (+0.0\%) & \cellcolor[gray]{0.5}0.1017 (+0.0\%) & \cellcolor[gray]{0.95}0.0999 & \cellcolor[gray]{0.5399}0.1071 (+2.0\%) & \cellcolor[gray]{0.5}0.1078 (+0.0\%) & \cellcolor[gray]{0.5}\textbf{0.1281} & \cellcolor[gray]{0.95}0.0649 (+47.6\%) & \cellcolor[gray]{0.505}0.1274 (+0.9\%) \\
\bottomrule
\end{tabular}
\end{adjustbox}
\end{table*}
 
\begin{figure}
    \centering
    \includegraphics[trim={0 0 0 0.2cm},clip,width=0.45\textwidth]{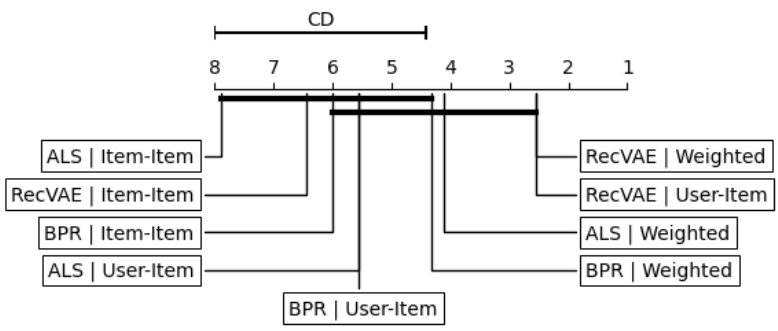}
    \caption{Visual representation of the Nemenyi test containing the critical difference (CD) and the models' average NDCG@10 rank with fine-tuning of the embeddings.}
    \Description[Visual representation of the Nemenyi statistical test]{Line with numbers ranging from 8 (left) to 1 (right). Nine boxes are positioned under the line, each one corresponding to a combination of embedding-based and recommender algorithms. Each box is connected to the line in the position of the average ranking of the algorithm, them being, in descending order from worst to best, ALS Item-Item, RecVAE Item-Item, BPR Item-Item, ALS User-Item, BPR User-Item, BPR Weighted, ALS Weighted, ReCVAE User-Item and RecVAE Weighted. Two strong bold lines representing the critical difference connects statistically similar algorithms. The leftmost line connects the bottom-6 algorithms, and the rightmost line connects the top-6.}
    \label{fig:nemenyi}
\end{figure}

To confirm the statistical significance of these results, we conducted a Friedman test~\cite{Demsar2006} on the NDCG@10 rankings from Table~\ref{tab:ndcg_finetuned}. The null hypothesis of no significant differences between the models was rejected with 95\% confidence ($X^2_r = 30.2815$), indicating that the models perform differently. We followed this with a post-hoc Nemenyi test (Figure~\ref{fig:nemenyi}), which showed that, with 90\% confidence and a critical difference (CD) of $3.59$, the weighted versions of RecVAE and ALS significantly outperform both item--item versions of ALS and BPR. Additionally, the ensemble methods consistently ranked equal to or higher than their original counterparts.

Although these findings do not provide conclusive evidence that our ensemble model is universally superior to the base algorithms, they demonstrate a consistent advantage in performance. Further investigation is warranted to explore the nuances of this improvement.

\section{Conclusions and future work}

In this paper, we presented a novel ensemble approach that integrates user--item and item--item recommendation strategies through a weighted similarity framework. Unlike traditional methods, our approach leverages pre-trained embeddings from embedding-based algorithms for both types of recommendations, effectively navigating the vector space learned by the embeddings to identify items similar to the user or other items they have previously consumed. 

We evaluated our ensemble model against the individual recommenders it is built upon, which comprised two efficient matrix factorization algorithms and a state-of-the-art neural network algorithm for learning embeddings. The results confirm that our model achieves competitive performance while maintaining high stability, excelling in scenarios that favor user--item or item--item recommendations. This stability and its computational efficiency demonstrate that our ensemble method offers reliable performance without requiring extensive fine-tuning or added complexity.

Our initial experiments utilized hyperparameters optimized for user--item recommendation. While fine-tuning embeddings specifically for the ensemble yields performance gains, these improvements are relatively modest compared to item--item recommendation methods. Therefore, customized fine-tuning can be safely omitted in many cases without significantly compromising the model's performance, which enhances its practical applicability.  

For future research, we propose investigating new embedding learning techniques to further refine the ensemble's effectiveness, incorporating SOTA embedding-based algorithms with different strategies for learning the representation. We also plan to analyze how the characteristics of different datasets impact the ensemble's behavior, finding scenarios where our model can benefit the most. Exploring distinct embeddings for each component recommender could enhance performance, although this may add to the model's complexity. Another promising direction is the development of methods to learn the similarity weights more intelligently and automatically, eliminating the need for manual fine-tuning and further streamlining the model's deployment in real-world scenarios.

\begin{acks}
This study was financed, in part, by the Brazilian Agencies CNPq (grant \#311867/2023-5), CAPES (Finance Code 88887.854357/2023-00) and FAPESP (Process Numbers \#2021/14591-7 and \#2023/00158-5).
\end{acks}

\balance
\bibliographystyle{ACM-Reference-Format}
\bibliography{arxiv_references} 

\end{document}